\documentclass[runningheads]{article}
\usepackage{graphicx}
\usepackage[affil-it]{authblk}
\usepackage{subcaption}
\usepackage[symbol]{footmisc}
\usepackage{amssymb}
\usepackage{amsmath}
\DeclareMathOperator*{\argmax}{arg\,max}
\DeclareMathOperator*{\argmin}{arg\,min}
\usepackage{wrapfig}
\usepackage{booktabs,tabularx}
\newcolumntype{x}{>{\raggedright\arraybackslash}X}
\usepackage{ctable}
\usepackage[colorlinks=true,
            linkcolor=magenta,
            urlcolor=gray,
            citecolor=cyan]{hyperref}

\usepackage{url}
\urldef{\mailsa}\path|{anjany.sekuboyina@tum.de}|

\begin{document}
\setlength{\belowcaptionskip}{-5pt}
\title{Probabilistic Point Cloud Reconstructions\\ for Vertebral Shape Analysis \thanks{Accepted at \textbf{M}edical \textbf{I}mage \textbf{C}omputing and \textbf{C}omputer-\textbf{A}ssisted \textbf{I}ntervention 2019.}}
%
%
\newcommand*\samethanks[1][\value{footnote}]{\footnotemark[#1]}
\author{Anjany Sekuboyina$^{1,2}$ }
\author{Markus Rempfler$^{3}$ }
\author{Alexander Valentinitsch$^{2}$ }
\author{Maximilian Loeffler$^{2}$ }
\author{Jan S. Kirschke$^{2}$\thanks{Joint supervising authors.}}
\author{Bjoern H. Menze$^{1}$\samethanks}
\affil{\small $^{1}$Department of Informatics, Technical University of Munich, Germany\\ $^{2}$Department of Neuroradiology, Klinikum rechts der Isar, Germany\\ $^{3}$Friedrich Miescher Institute for Biomedical Research, Switzerland\\
\mailsa}
\date{}

%
%

%
\maketitle              
\begin{abstract}
We propose an auto-encoding network architecture for point clouds (PC) capable of extracting shape signatures without supervision. Building on this, we (i) design a loss function capable of modelling data variance on PCs which are unstructured, and (ii) regularise the latent space as in a variational auto-encoder, both of which increase the auto-encoders' descriptive capacity while making them probabilistic. Evaluating the reconstruction quality of our architectures, we employ them for detecting vertebral fractures without any supervision. By learning to efficiently reconstruct only healthy vertebrae, fractures are detected as anomalous reconstructions. Evaluating on a dataset containing $\sim$1500 vertebrae, we achieve area-under-ROC curve of $>$75\%, without using intensity-based features.     

\end{abstract}

\section{Introduction}
One of the consequences of the numerous algorithms proposed for segmenting organs, tissues, the spine etc. involves analysing their anatomical shapes, eventually contributing towards population studies \cite{Ingalhalikar14}, disease characterisation \cite{shakeri16}, survival analysis \cite{isensee18}, etc. Employing convolutional neural networks (CNN) for this task involves processing voxelised data due to its Euclidean nature. Such voluminous representation, however, is inefficient, especially when the masks are binary and the \emph{shape information} corresponds to its surface profile. Alternatively, surface meshes (a collection of vertices, edges, and faces) or active contours could be used. Since the data is no longer Euclidean, a conventional CNN is unusable. Graph convolutional networks (GCN) \cite{bronstein17} were thus developed by redefining the notion of `neighbourhood' and `convolution' for meshes and graphs. However, if the number of nodes is high, GCNs (esp. spectral) become bulky. Moreover, each mesh is treated as a domain, making mesh registration a requisite. 

An alternative surface representation is a set of 3D points in space, referred to as the \textbf{point clouds} (PC). A PC represents the surface just with a set of $N$ vertices, thus avoiding both the cubic-complexity of voxel-based representations and the $N \times N$ dimensional, sparse, adjacency matrix of meshes. However, despite their representational effectiveness, PCs are permutation invariant and do not describe data on a structured grid, preventing the usage of standard convolution. To this end, we work with an architecture capable of processing PCs (\emph{point-net}, \cite{charlesPN}), and design a network capable of reconstructing PCs thereby extracting shape signatures in an unsupervised manner.

\subsubsection{Uncertainty and latent space modelling} Unlike supervised learning on PCs \cite{benjamin18}, we set out to obtain shape signatures from PCs without supervision, building towards a relatively less explored topic of \emph{auto-encoding} point clouds. This involves mapping the PC to a latent vector and reconstructing it back. Since the PCs are unordered, PC-specific reconstruction losses replace traditional ones \cite{charlesPNAE,yang18_foldingnet}. Extending auto-encoders (AE) based on such a loss, we propose to improve its representational capacity by regularising the latent space to make it compact and by modelling the variance that exists in a PC population.  We claim that this results in learning improved shape signatures, validating the claim by employing the extracted features for unsupervised vertebral fracture detection.   

\begin{wrapfigure}{r}{0.46\textwidth} 
\vspace{-20pt}
 \centering
    \includegraphics[width=0.45\textwidth]{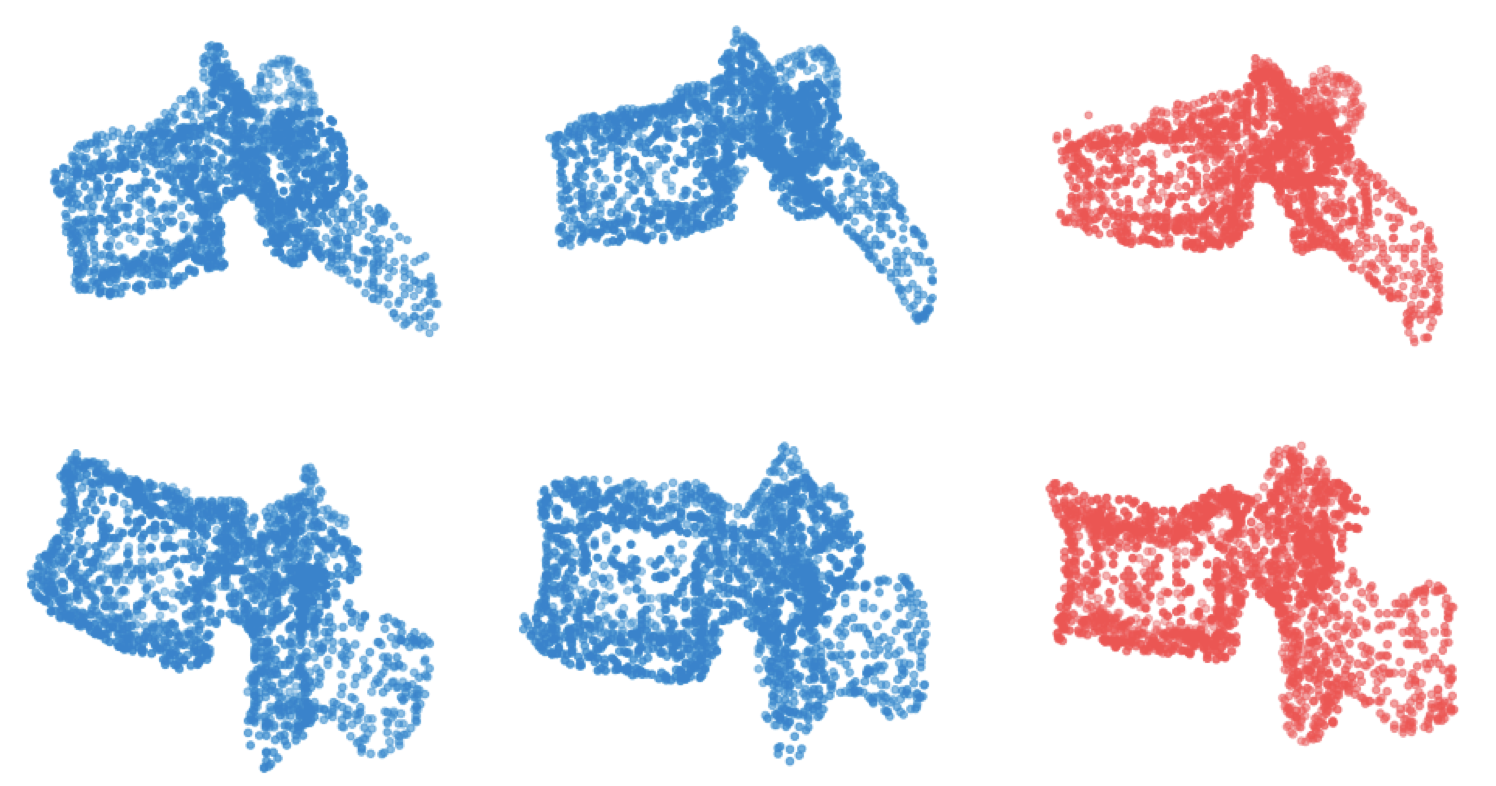}
  \caption{\scriptsize \textbf{Variation among vertebral shapes}: Compare the higher variation between healthy (blue) vertebrae of different classes (T3, top and L1, bottom) w.r.t the relatively lower variation within-class between fractured (red) and healthy vertebrae.}
      \label{figure:motivation}
\end{wrapfigure}

\subsubsection{Vertebral fracture detection} There exists an inherent shape variation in vertebral shapes within the spine of a single patient (e.g. cervical--thoracic--lumbar) along with a natural variation in a vertebra's shape in a population (e.g. L1 across patients, cf. Fig. \ref{figure:motivation}). Additionally, osteoporotic fractures start without significant shape change and progress into a vertebral collapse. Hence, fracture detection in vertebrae is non-trivial. Added to this, limited availability of fractured vertebrae makes the learning of supervised classifiers non-trivial. In literature, several classification systems exist mainly based on vertebral height measurement \cite{baum14} or analysing sub-regions of the spine in sagittal slices \cite{tomita18}. However, an explicit shape-based approach seems absent. Evaluating the representational ability of the proposed AE architectures, we seek to analyse vertebral shapes and eventually detect vertebral fractures using the extracted latent shape features.

\subsubsection{Our contribution}  
Summarising the contributions of this work: (1) We build on existing point-net-based architectures to propose a point-cloud auto encoder ($p$AE). (2) Reinforcing this architecture, we incorporate latent space modelling and a more challenging uncertainty quantification. (3) We present a comprehensive analysis of the reconstruction capabilities of our $p$AEs by investigating their utility in detecting vertebral fractures. We work with an in-house, clinical dataset ($\sim$1500 vertebrae) achieving an area-under-curve (AUC) of $>$75\% in detecting fractures, even without employing texture or intensity-based features.

\section{Methodology}

We present this section in two stages: First, we introduce the notation used in this work and describe a point-net-based architecture capable of efficiently auto-encoding point clouds. Second, we build on this architecture to model the natural variance in vertebrae while regularising the latent space.  

\subsection{Auto-encoding point clouds} 
Given accurate voxel-wise segmentation of a vertebra, a \emph{point cloud} (PC) can be extracted as a set of $N$ points represented by $X = \{p_i\}_{i=0}^N$, where $p_i$ represents a point by its 3D coordinate $(x_i, y_i, z_i)$. Additionally, $p_i$ could also represent other point specific features such as normal, radius of curvature etc. So, each vertebra is represented by a PC of dimension $N \times m$ (in this work, $N=2048$ vertices and $m=3$ coodinates, with the vertices randomly subsampled from a higher resolution mesh). Recall the lack of a regular coordinate space associated with the point cloud and that any permutation of these $N$ points represents the same point cloud. This requires incorporation of a unique variant of deep networks for processing PCs.\\

\begin{figure*}[t!]
\centering
       \includegraphics[width=0.85\textwidth]{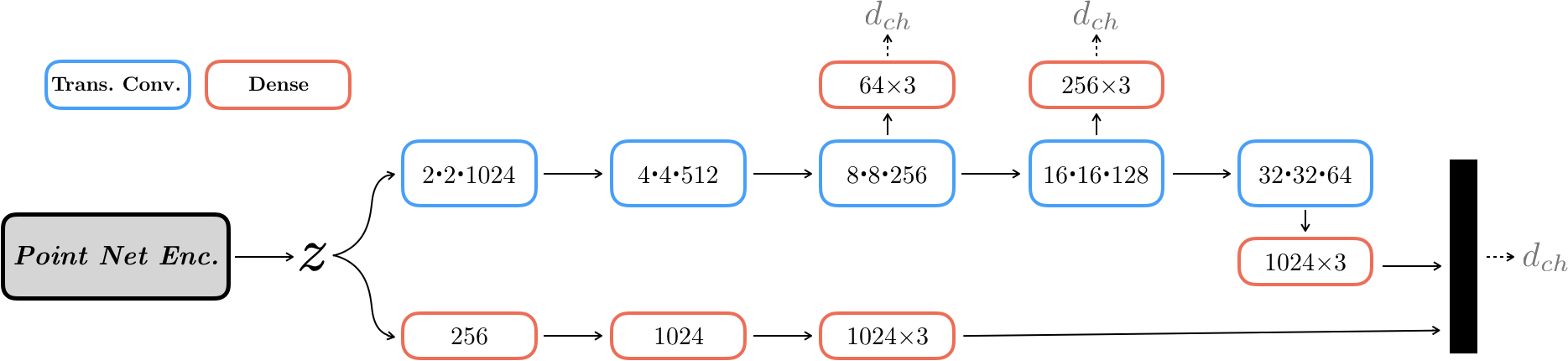}
       \caption{\footnotesize \textbf{Point Cloud Auto-encoder ($p$AE)}: Architectural details of decoding path constructing a point cloud from a latent vector. Top arm is convolutional while bottom arm is fully-connected. Transposed convolution ($-\cdot-\cdot\text{channels}$) have a stride of 2. Since encoder is an adapted \emph{point-net}\cite{charlesPN}, we detail its architecture in the appendix.}
      \label{figure:point_ae}
\end{figure*}

\noindent
\textbf{Architecture.}
An AE consists of an encoder mapping the PC to the latent vector and a decoder reconstructing the PC back from this latent vector, i.e $X \mapsto z \mapsto X$. As the encoder, we employ a variant of the point-net architecture \cite{charlesPN}. The latent vector, $z$, respects the permutation invariance of the PC and represents its shape signature. As a decoder, taking cues from \cite{charlesPNAE}, we construct a combination of an up-convolutional and dense branches taking $z$ as input and predicting $\hat{X}$, the reconstructed $X$. The convolutional path, owing to its neighbourhood processing, models the `average' regions, while the dense path reconstructs the finer structures. This combination of the point-net and the decoder forms our point cloud auto-encoding ($p$AE, or interchangeably AE) architecture as illustrated in Fig. \ref{figure:point_ae}.\\

\noindent 
\textbf{Loss.}
Reconstructing point clouds requires comparing the predicted PC with the actual PC to back-propogate the loss during training. However, owing to the unordered nature of PCs, usual regression losses cannot be employed. Two prominent candidates for such a task are the Chamfer distance and the Earth Mover (EM) distance \cite{charlesPNAE}. We observed that minimising EM distance ignores the natural variation in shapes (e.g. the processes of the vertebrae) and reconstructs only a mean representation (e.g. the vertebral body), as validated in \cite{charlesPNAE}. Since we intend to model the natural variance in the data, using EM distance is undesirable in our case. We thus employ the Chamfer distance computed as:
\begin{equation}
 d_{ch}(X, \hat{X}) = \mathcal{L}_{ae} = \sum_{p \in X} \min_{\hat{p} \in \hat{X}} || p - \hat{p} ||_2^2 +  \sum_{\hat{p} \in \hat{X}} \min_{p \in X} || p - \hat{p} ||_2^2.
 \label{eq:chamfer}
 \end{equation}
In essence, $d_{ch}$ is the distance between a point in $X$ and its nearest neighbour in $\hat{X}$ and vice versa.

\subsection{Probabilistic reconstruction}
From a generative modelling perspective, an AE can be seen to predict the parameters of Gaussian distribution imposed on $X$, i.e. $p_{\Theta}(X)$ =  $\mathcal{N}(X | \hat{X}, \hat{\mathrm{\Sigma}})$, parameterised by the weights of the AE denoted by $\Theta$. Determining the distribution parameters, viz. optimising for the AE weights, now involves maximising the log-likelihood of $X$, resulting in:
 \begin{equation}
\Theta^* = \argmax_\Theta \log p_\Theta(X) = \argmin_\Theta \frac{1}{2}(X - \hat{X})^T\hat{\mathrm{\Sigma}}^{-1}(X - \hat{X}) + \frac{1}{2}\log |\hat{\mathrm{\Sigma}}|.
 \label{eq:basis}
 \end{equation}
This perspective towards auto-encoding enables us to extend the $p$AE to encompass the data variance ($\hat{\mathrm{\Sigma}}$) while modelling the latent space, as described in following sections. It is important to note that the difference $X-\hat{X}$ is not well defined for point clouds, requiring us to opt for alternatives.

\begin{figure*}[t!]
\centering
\begin{minipage}[c]{0.49\textwidth}
       \includegraphics[width=\textwidth]{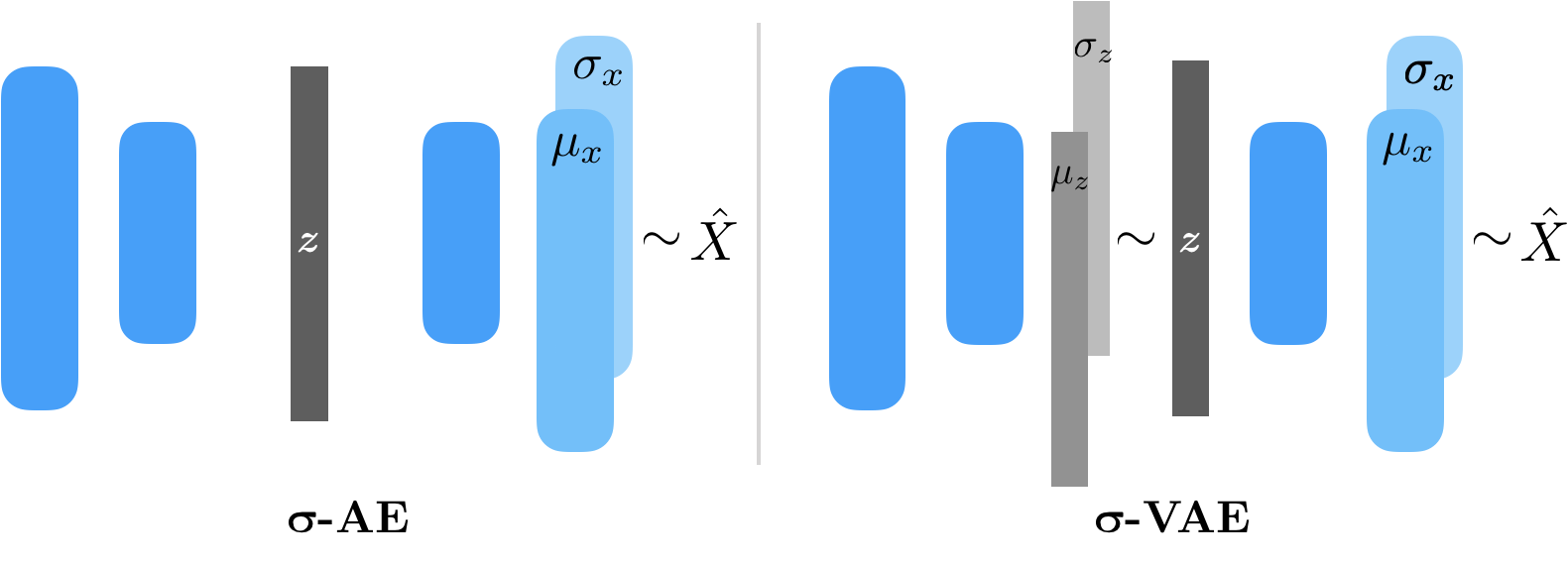}
\end{minipage}\hfill
\begin{minipage}[c]{0.48\textwidth}
       \caption{\footnotesize \textbf{Probabilistic reconstruction architectures}: $\sim$ indicates a sampling operation. Since a point's variance has a smaller scale compared to its mean, the variance is predicted using a softplus activation (added with $\epsilon = 10^{-6}$ for stabilising divisions) and uses a layer parallel to the one predicting the mean.}
      \label{figure:archs}
\end{minipage}
\end{figure*}

Assuming $\mathrm{\Sigma} = \mathbb{I}$,  implying an independence among the elements of $X$ and an element-wise unit variance, results in the familiar mean squared error (MSE), $\mathcal{L} = ||X - \hat{X}||^2$. Based on the parallels between MSE and the Chamfer distance (Eq. \ref{eq:chamfer}), we design $\sigma$-AE and $\sigma$-VAE, as illustrated in Fig \ref{figure:archs}.


\subsubsection{$\sigma$-AE.} The assumption of unit covariance, as in AE, is inherently restrictive. However, modelling an unconstrained covariance matrix is infeasible due to quadratic complexity. A practical compromise is the independence assumption. Thus, representing covariance as, $\mathrm{\Sigma} = diag\{\hat{\sigma}_{p_1}^2, \dots, \hat{\sigma}_{p_i}^2, \dots, \hat{\sigma}_{p_N}^2\}$, where $\hat{\sigma}_{p_i}^2$ denotes the variance corresponding to $p_i$, eq. (\ref{eq:basis}) morphs to a loss function as:
  \begin{equation}
 \mathcal{L} =  \sum_{\hat{p} \in \hat{X}} \sigma_{\hat{p}}^{-2} ||p_i - \hat{p_i}||^2 + \log \sigma_{\hat{p}}^2
 \label{eq:sigma_ae}
 \end{equation} 
 This optimisation models the aleoteric uncertainty \cite{kendall17}. Eq. \ref{eq:sigma_ae} is an attenuated MSE, where a high variance associated to a point down-weighs its contribution to the loss. However, due to the lack of a reference grid in the point cloud space, the notion of uncertainty being associated to a data point (eg. pixel, spatial location etc.) is absent. We propose to associate the notion of variance to every point, $\hat{p_i}$. This results in the variance-modelling Chamfer distance:
 \begin{equation}
 \mathcal{L}_{\sigma ae} = \sum_{p \in X} \min_{\hat{p} \in \hat{X}} \sigma_{\hat{p}}^{-2}|| p - \hat{p} ||_2^2 +  \sum_{\hat{p} \in \hat{X}} \sigma_{\hat{p}}^{-2} \min_{p \in X} || p - \hat{p} ||_2^2 + \log \sigma_{\hat{p}}^2
 \label{eq:sigma_ae_ch}
 \end{equation}

Observe the slight abuse of notation in Eq.~\ref{eq:sigma_ae_ch}, wherein the variance at a predicted point, $\sigma_{\hat{p}}$, actually represents the variance of the coordinate elements of $p$, i.e $\{\sigma_{\hat{x}}, \sigma_{\hat{y}}, \sigma_{\hat{z}}\}$. Current notation is chosen to avoid clutter.

\subsubsection{Variational and $\sigma$-Variational AE.} An alternative approach for modelling $p(X)$ involves modelling its dependency over a latent variable $z$, which is distributed according to a known prior $p(z)$. A variational auto-encoder (VAE) operates on these principles and involves maximising a lower bound on the log-evidence (referred to as ELBO) of the data described as below:

 \begin{equation}
 \log p(X) \geq \mathbb{E}_{z\sim q_\phi(z|X)}\big[\log p_\theta(X|z)\big] - \textrm{KL}\big[q_\phi(z|X)~||~p_\theta (z)\big],
 \label{eq:vae}
 \end{equation}
where $q_\phi(z|x)$ is the approximate posterior of $z$ learnt by the encoder and parameterised by $\phi$. $p_\theta(X|z)$ is the data likelihood modelled by the decoder and parameterised by $\theta$. $p_\theta(z)$ is the prior on $z$. 

Maximising ELBO is equivalent to maximising the log-likelihood of $X$ while minimising the Kullback-Leibler divergence between the approximate and true prior. Representing the combination as $\mathcal{L}_{rec} + \beta \mathcal{L}_{KL}$, where $\mathcal{L}_{rec}$ is the reconstruction loss seen is earlier sections. $\beta$ is a scaling factor weighing the contribution of the two losses appropriately. Standard practice assigns Gaussian distributions for $q_\phi(z|x) \sim \mathcal{N}(z | \boldsymbol{\mu}_z, \boldsymbol{\sigma}_z)$ and $p(z) \sim \mathcal{N}(z | \mathbf{0}, \mathbf{1})$ (cf. Fig. \ref{figure:archs}). Thus, $\mathcal{L}_{KL}$ models the latent space to follow a Gaussian distribution inline with the prior. Incorporating this into the point cloud domain, results in an objective function for a PC-based VAE (or $\sigma$-VAE) as $\mathcal{L}_{vae} = \mathcal{L}_{ae/\sigma ae} + \beta \mathcal{L}_{KL}$. Thus, $\sigma$-VAE acts as a AE capable of modelling the data variance while regularising the latent space. The prior on the latent space also imparts point cloud generation capabilities to $\sigma$-VAE.

 \subsection{Detecting fractures as anomalies}
 \label{methodology:frac_detection}
Examining the descriptive ability of our $p$AE architectures in auto-encoding PCs, we utilise them for detecting vertebral fractures. Assuming the AE is trained only on `normal' patterns,  a fracture can be detected as an `anomaly' based on its `position' in latent space. We inspect two measures for this purpose: 
\begin{enumerate}
\item
Reconstruction error or Chamfer distance: AEs trained on healthy samples fail to accurately reconstruct anomalous ones, resulting in a high $d_{ch}$.
\item
Reconstruction probability or likelihood  \cite{an15}: Expected likelihood $\mathbb{E}\big[p_\Theta(X)\big]$ of an input can be computed for $\sigma-$ architectures (cf. Eq. \ref{eq:basis}). For any input PC, $X_{in}$, it is computed by $\mathcal{N}(X_{in} | \mu_\Theta,\Sigma_\Theta)$ with the predicted mean and variances. We expect fractured vertebrae to be less \emph{likely} than healthy ones.
\end{enumerate}

Intuitively, relying on the reconstruction error or likelihood for detecting anomalies requires the learnt `healthy' latent space to be representative. Both $\sigma$-AE and the VAE work towards this objective. In $\sigma$-AE, predictive variance down-weighs the loss due to highly uncertain points in the PC. This suppresses the interference due to natural variation in the vertebral PCs. On the other hand, VAE acts directly on the latent space by modelling the encoding uncertainty ($X\mapsto z$). The $\sigma$-VAE encompasses both these features.

\subsubsection{Inference.} A given vertebral PC is reconstructed and the reconstruction error and (or) likelihood are computed. This vertebra is said to be fractured if the reconstruction error is greater than a threshold, $T_{rec}$, or its likelihood is lesser than a threshold, $T_{l}$. $T_{rec}$ and $T_{l}$ and determined on the validation set.

%
%
%

\section{Experiments \& Discussion}
We present this section in two parts: first, we explore the auto-encoding, variance modelling, and generative capabilities of our AE networks. Second, we deploy these architecture to detect vertebral fractures without supervision.

\noindent
\emph{Data preparation}: We evaluate our architecture on an in-house dataset with accurate voxel-level segmentations converted into PCs. The dataset consists of 1525 healthy and 155 fractured vertebrae, denoted as ($1525H + 155F$) vertebrae. Since we intend to learn the distribution of healthy vertebrae, we do not use any fractured vertebrae during training. The validation and test sets consists of ($50H + 55F$) and ($100H+100F$) vertebrae, respectively. For the supervised baselines, the train set needs to contain fractured vertebrae. Thus, validation and test sets were altered to contain ($50H + 55F$) and ($55H + 55F$) vertebrae. 

\noindent
\emph{Training}: The architecture of the encoder and the decoder is similar across all architectures (cf. Fig \ref{figure:archs}) except for the layers predicting variance. PCs are augmented online by perturbing the points with Gaussian noise and random rotations ($\pm 15\deg$). Finally, the PCs are median-centred to origin and normalised to have the same surface area. The networks are trained until convergence using an Adam optimiser with an initial learning rate of $5 \times 10^{-4}$. Specific to the VAE, we use KL-annealing by increasing $\beta$ from 0 to 0.1.

\begin{figure*}[t!]
\centering
       \includegraphics[width=0.95\textwidth]{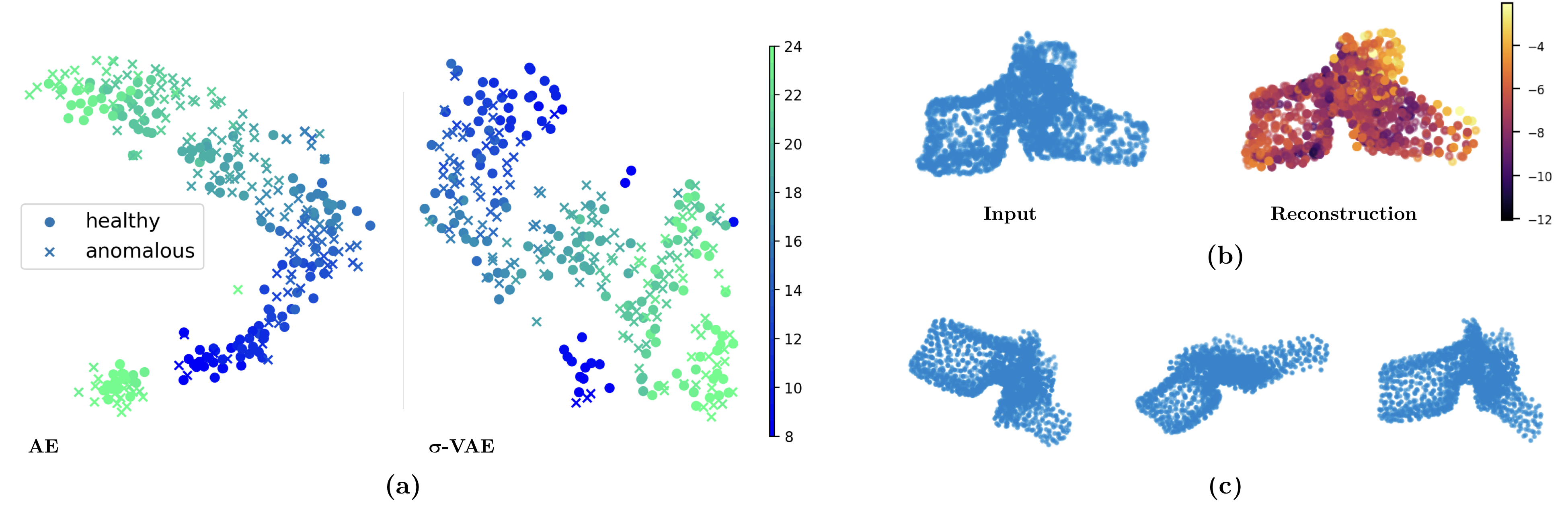}
       \caption{\footnotesize \textbf{Characteristics of $\sigma$-VAE}: (a) Comparison of TSNE embeddings of simple $p$AE with $\sigma$-VAE. Observe transition in clusters being inline with vertebral indices. Note that embedding becomes compact for a VAE. (b) A PC and its reconstruction coloured with $\log(\sigma^2)$ of every point. Observe high variance in vertebral processes. (c) Example generations from decoder with $z\sim\mathcal{N}(\mathbf{0}, \mathbf{1})$.}
      \label{figure:qual}
\end{figure*}

\subsubsection{Qualitative evaluation of AE architectures.} 
We investigate if meaningful shape features can be learnt without supervision. Validating this, in Fig. \ref{figure:qual}a, we plot a TSNE embedding of the test set latent vectors learnt by a naive $p$AE and $\sigma$-VAE trained only on healthy vertebrae. Observe the clusters formed based on the vertebral index and the transition between the indices. This corresponds to the natural variation of vertebral shapes in a human spine. Indicating the fractured vertebrae in the embedding, we highlight their degree of similarity with the healthy counterparts. Also, observe that embedding is more regularised representing a Gaussian in case of $\sigma$-VAE, indicating the continuity of the learnt latent space. Fig. \ref{figure:qual}b shows the predictive variance modelled by the $\sigma$-VAE. Posterior elements of a vertebrae are the most varying among population. Observe this being captured by the variance in the vertebral process regions. Lastly, illustrating $\sigma$-VAE's generative capabilities, Fig. \ref{figure:qual}c shows vertebral PC samples generated by sampling the latent vector, $z\sim \mathcal{N}(\mathbf{0}, \mathbf{1})$.

\subsubsection{Vertebral fracture detection.}
Evaluating the reconstruction quality of our $p$AE architectures, we employ them to detect fractures as anomalies. As baselines, we choose two supervised approaches: (1) point-net (PN), the encoding part in our $p$AE architectures, cast as a binary classifier and (2) the same point-net trained with median frequency balancing the classes (ref. as PN$_{bal}$) to accentuate the loss from minority fractured class. We report their performance in Table \ref{table:pr}, over 3-fold cross-validation while retaining the ratio of healthy to fractured vertebrae in the data splits. Frequency balancing improves the F1 score significantly, albeit not at the level of the proposed anomaly detection schemes.


\begin{figure*}[t!]
\centering
\begin{minipage}[c]{0.64\textwidth}
       \includegraphics[width=\textwidth]{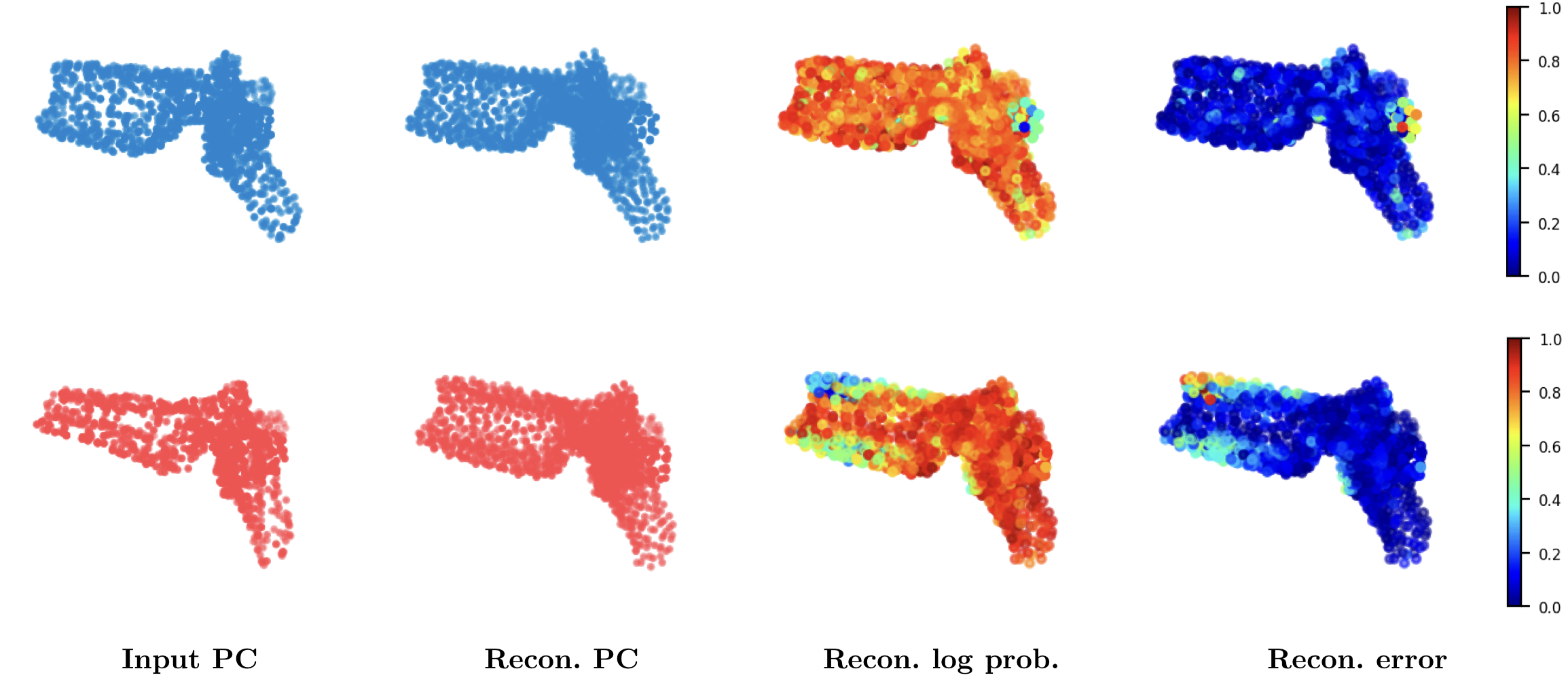}
\end{minipage}\hfill
\begin{minipage}[c]{0.34\textwidth}
       \caption{\footnotesize \textbf{Reconstructions}: Healthy (top) and fractured (bottom) vertebral PCs. Observe $p$AE's `healthy' reconstruction of a fracture. Errors and log-probabilities are normalised to [0,1] within PC for visualisation, but anomaly detection works on un-normalised values.}
      \label{figure:recons}
\end{minipage}
\end{figure*}

\begin{table*}[t!]
 \caption{\footnotesize Performance comparison of unsupervised and supervised fracture detection approaches. Measures: Precision (P), Recall (R), F1-score, and Area-Under-ROC Curve (AUC) computed by varying thresholds on recon. error and recon. log-probabilities. Since supervised models have no threshold selection, AUC is not reported.}
 \scriptsize
  \centering
   \setlength{\tabcolsep}{0.3em}
\begin{tabular}{c | cc | cccc | cc}
	\multicolumn{3}{c}{} & \multicolumn{4}{c}{\emph{recon. error}} & \multicolumn{2}{c}{\emph{recon. log-likelihood}}\\
	\toprule
	\rule{0pt}{2ex} Measures & PN & PN$_{bal}$ & AE & VAE & $\sigma$-AE & $\sigma$-VAE & $\sigma$-AE & $\sigma$-VAE\\ [0.25ex] 
 	\midrule
	 \rule{0pt}{2ex}$P$ & 100$\pm$0.0 & 68.6$\pm$3.4 & 57.6$\pm$4.1 & 61.1$\pm$1.9  & 67.1$\pm$6.5 & \textbf{68.4$\pm$3.3} & \textbf{62.3$\pm$4.3}  & 61.6$\pm$1.4\\
	\rule{0pt}{2ex}$R$ & 13.9$\pm$3.1 & 57.6$\pm$7.5 & \textbf{85.0$\pm$9.8} & 79.0$\pm$3.6  &74.3$\pm$4.0& 71.7$\pm$4.1 & 72.7$\pm$6.1 & \textbf{79.7$\pm$2.5}\\
	\rule{0pt}{2ex}$F1$ & 24.7$\pm$4.7 & 62.5$\pm$5.8 & 68.0 $\pm$0.9 & 68.5$\pm$1.7  & 67.5$\pm$5.1& \textbf{69.6$\pm$1.2} & 66.7$\pm$1.3 & \textbf{69.5$\pm$0.6}\\
	\rule{0pt}{2ex} AUC & n.a & n.a & 70.8 $\pm$2.2 & 74.8$\pm$3.0 &\textbf{ 75.9$\pm$2.0} & \textbf{75.9$\pm$1.5} & 70.2$\pm$2.2 & \textbf{73.8$\pm$2.0}\\
 	\midrule
\end{tabular}
\label{table:pr}
\end{table*}

\noindent
\emph{Reconstruction for fracture detection}: When detecting fractures based on reconstruction error ($d_{ch}$), we observe that a naive $p$AE already out-performs the supervised classifiers (cf. Table \ref{table:pr}). On top of this, we see that latent space modelling and variance modelling individually offer an improvement in F1-scores while increasing the AUC, indicating a stable detection of fractures. The performance of both $\sigma$-AE and $\sigma$-VAE is similar indicating the role of loss attenuation. However, the advantage of explicitly regularising the latent space for $\sigma$-VAE can be seen in likelihood-based anomaly detection, where $\sigma$-VAE outperforms $\sigma$-AE. Fig. \ref{figure:recons} compares a reconstruction of a healthy and fractured vertebrae of the same vertebral level. Note the high reconstruction error and a low log-likelihood spatially corresponding to the deformity due to fracture.

\section{Conclusions}
We presented point-cloud-based auto-encoding architectures for extracting descriptive shape features. Improving their description, we incorporated variance and latent space-modelling capability using specially defined PC specific losses. The former captures the natural variance in the data while the latter regularises the latent space to be continuous. Deploying these networks for the task of unsupervised fracture detection, we achieved an AUC of 76\% without using any intensity or textural features. Future work will combine the extracted shape signatures with textural features e.g. bone density and trabecular texture of vertebrae to perform fracture-grade classification.\\

\noindent
\textbf{Acknowledgements.} This work is supported by the European Research Council (ERC) under the European Union's `Horizon 2020' research \& innovation programme (GA637164--iBack--ERC--2014--STG). The Quadro P5000 used for this work was donated by NVIDIA Corporation.

\bibliographystyle{plain}
\bibliography{bibliography}

\newpage
\section*{Appendix}
As supplementary content, we present: (1) A detailed description of the complete point-cloud auto-encoder, including the encoder architecture adapted from \emph{point net} \cite{pn} (cf. Fig.~\ref{figure:pn_ae}), (2) additional illustrations of point-wise data uncertainty modelled by the proposed $\sigma$-VAE (cf. Fig.~\ref{figure:sigmas}), and (3) Further qualitative results comparing probabilistic reconstructions of  healthy and anomalous or fractured vertebrae, along with point-wise Chamfer distance and log-probability between the input and its reconstruction (cf. Fig.~\ref{figure:recons}). 

\setcounter{figure}{5} 
\begin{figure*}[h]
\centering
       \includegraphics[width=0.85\textwidth]{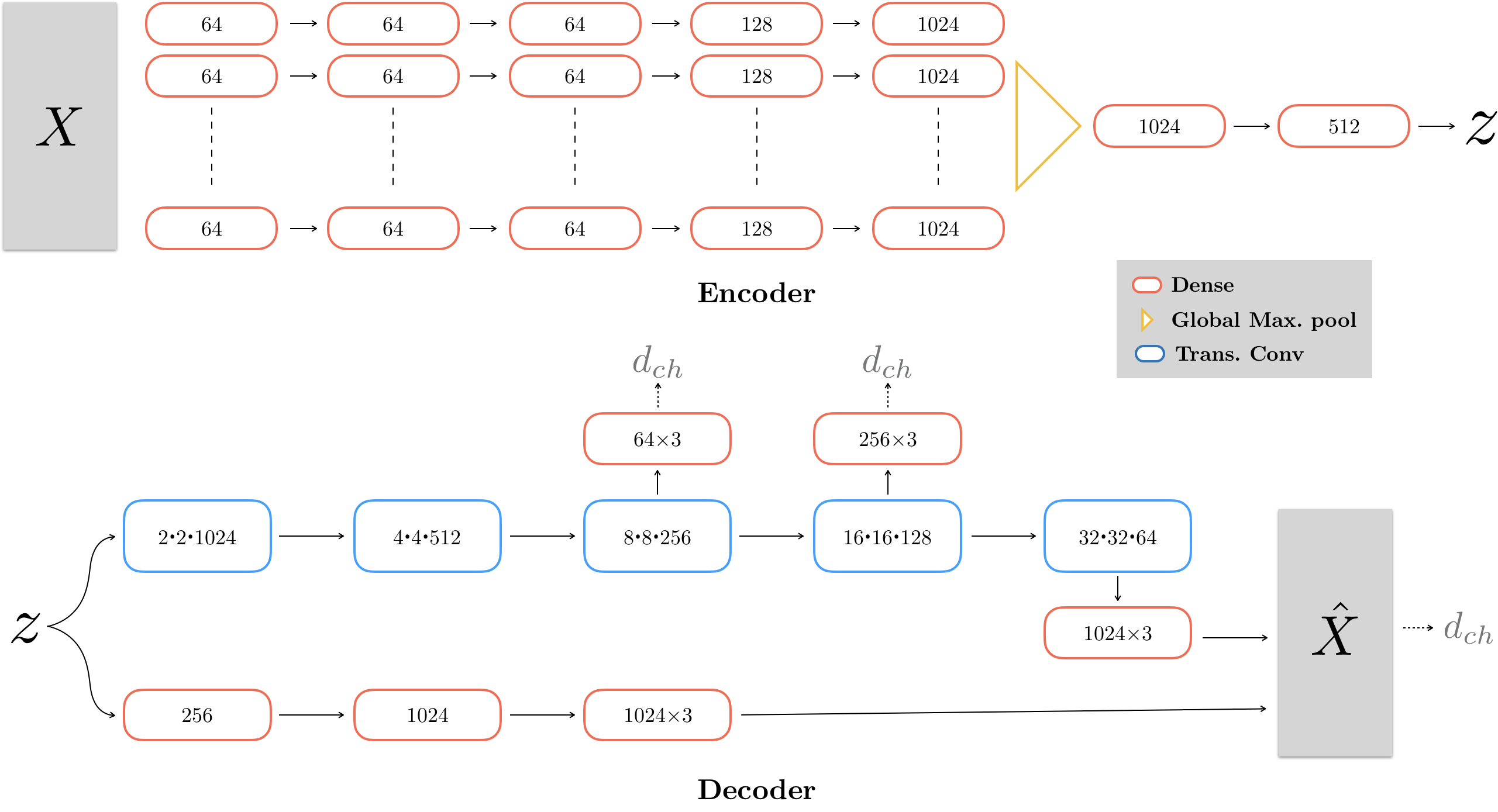}
       \caption{\footnotesize \textbf{Architecture for $p$AE}: Architectural details of encoding and decoding paths of the $p$AE. Note that every layer (except the last, in encoder and in the decoder) is followed by batch normalisation and leaky ReLU. The values in the dense layers indicate the number of nodes while the values in the transposed convolution layers ($-\cdot-\cdot\text{channels}$) indicate the size of the resulting feature map. For example, the first transposed convolution layer, $z$ is reshaped to $1\cdot1\cdot64$ and up-convolved to $2\cdot2\cdot1024$.}
      \label{figure:pn_ae}
\end{figure*}

\begin{figure*}[h]
\centering
       \includegraphics[width=0.85\textwidth]{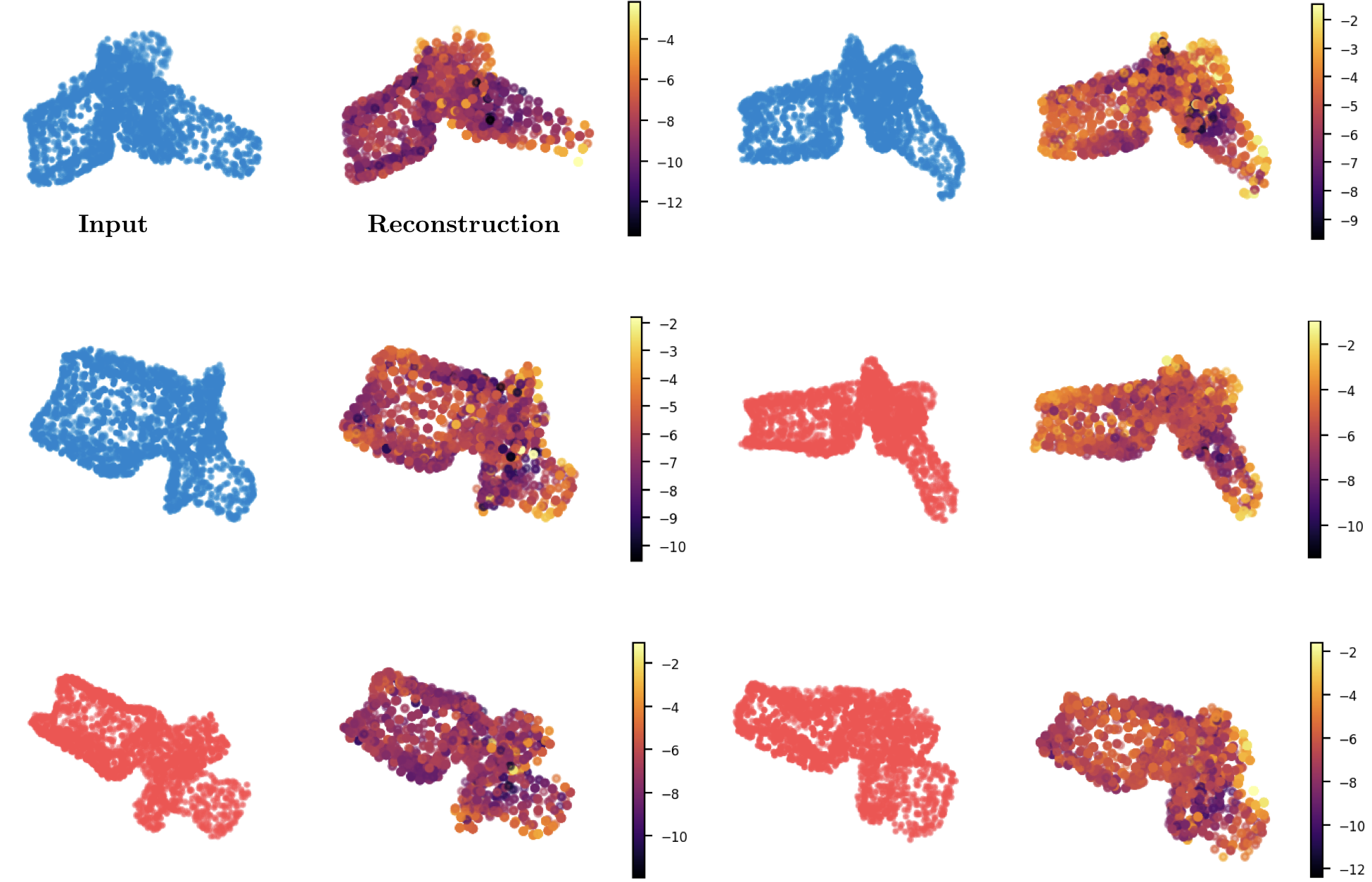}
       \caption{\footnotesize \textbf{Variance modelling} by the proposed $\sigma$-VAE for healthy (blue) and fractured (red) cases. Observe a higher variance in the vertebral processes representing the naturally occurring shape variance in a population. Note the lack of high uncertainty values for fractured vertebrae, in accordance with aleoteric uncertainty's property of capturing only data variance \cite{kendall}. Hence, predictive variance cannot be used as a means to detect fractures. However, it can reliably be employed as an attenuation factor for improving reconstruction or for computing the reconstruction probability (cf. Fig.~\ref{figure:recons}), thereby enabling fracture detection.}
\label{figure:sigmas}
\end{figure*}

\begin{figure*}[t!]
\centering
       \includegraphics[width=0.85\textwidth]{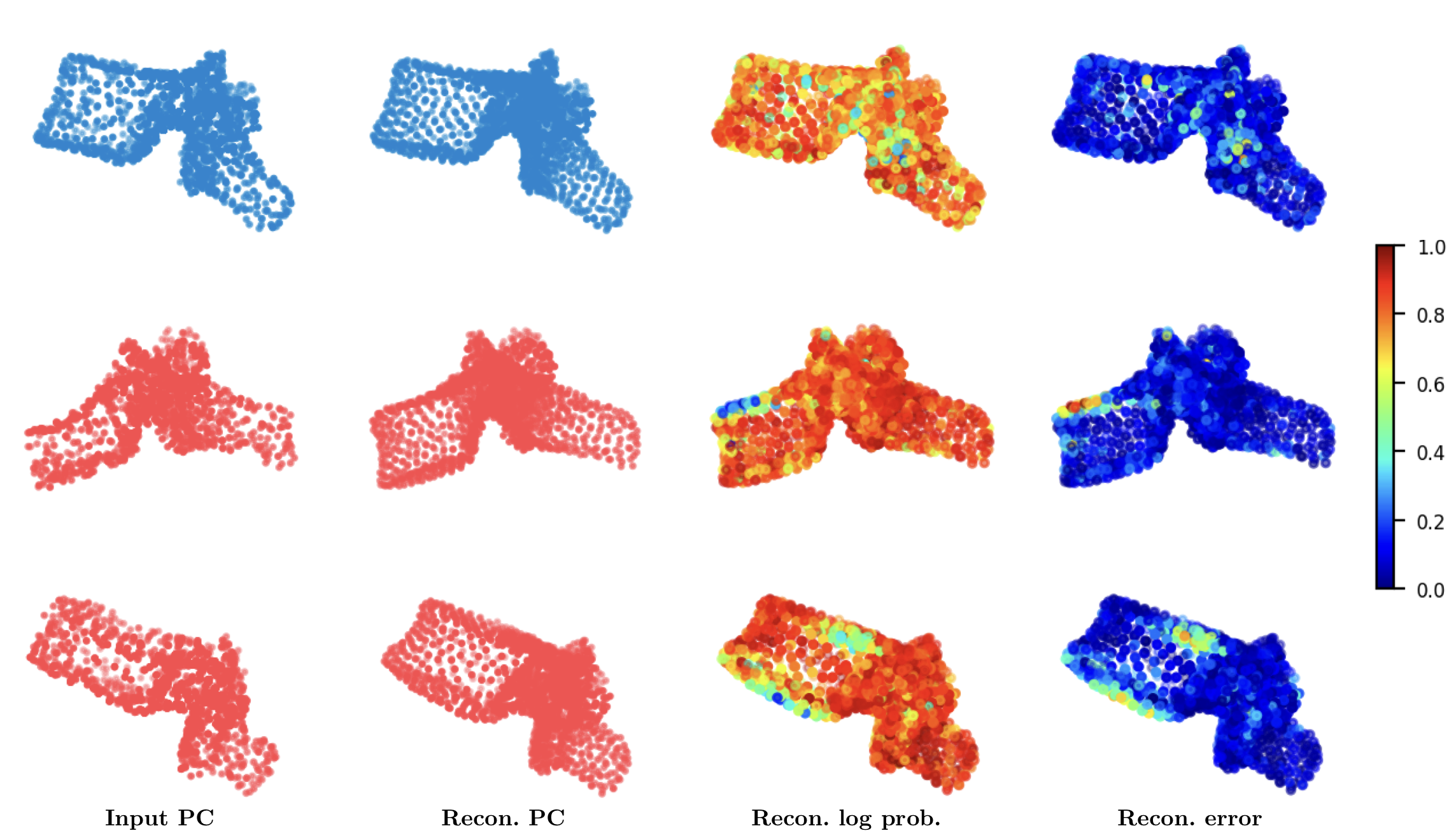}
       \caption{\footnotesize \textbf{Probabilistic reconstructions} of healthy and fractured vertebrae. Alongside spatial localisation of fractures, compare the dynamic range of the reconstruction log probability between healthy and fractured cases when normalised to [0,1]. Reconstruction probabilities are relatively uniformly-spread in the healthy case and are pushed to the extremes for a fractured one. This indicates a higher dynamic range in the unnormalised values while reconstructing fractured vertebrae. Thus, a low reconstruction probability (or a high reconstruction error) does indicate an outlier or a fracture.}
      \label{figure:recons}
\end{figure*}

\end{document}